\documentclass{ptephy_v1}%%%%%% to generate preprint number with ptep logo

\usepackage{amsmath}
\usepackage{amsthm}
\usepackage{graphics}
\usepackage{natbib}

\begin{document}

\title{Scintillation Balloon for Neutrinoless Double-Beta Decay Search with Liquid Scintillator Detectors}

%%%%%%%%%%%%%%%%%%%%%%%%%%%%%%%%%%%%%%%%%%%%%%%%%%%%%%%%%%%%%%%%%%%%%%%%%%%%%%%%%%%%%%
\author{S.~Obara*\thanks{Current address; {\it Department of Physics, Kyoto University, Kyoto 606-8502, Japan}}}
\author{Y.~Gando}
\author{K.~Ishidoshiro}

\affil{Research Center for Neutrino Science, Tohoku University, Sendai 980-8578, Japan}

\affil{\email{obara@awa.tohoku.ac.jp}}

%%%%%%%%%%%%%%%%%%%%%%%%%%%%%%%%%%%%%%%%%%%%%%%%%%%%%%%%%%%%%%%%%%%%%%%%%%%%%%%%%%%%%%
\begin{abstract}
Environmental radioactivity is a dominant background for rare decay search experiments, and it is difficult to completely remove such an impurity from detector vessels.
We propose a scintillation balloon as the active vessel of a liquid scintillator in order to identify this undesirable radioactivity.
According to our feasibility studies, the scintillation balloon enables the bismuth--polonium sequential decay to be tagged with a 99.7\% efficiency, assuming a KamLAND (Kamioka Liquid scintillator AntiNeutrino Detector)-type liquid scintillator detector. 
This tagging of sequential decay using alpha-ray from the polonium improves the sensitivity to neutrinoless double-beta decay with rejecting beta-ray background from the bismuth.
\end{abstract}

%%%%%%%%%%%%%%%%%%%%%%%%%%%%%%%%%%%%%%%%%%%%%%%%%%%%%%%%%%%%%%%%%%%%%%%%%%%%%%%%%%%%%%
\subjectindex{double-beta decay, liquid scintillator, polyethylene naphthalate, Majorana neutrino}

\maketitle

%%%%%%%%%%%%%%%%%%%%%%%%%%%%%%%%%%%%%%%%%%%%%%%%%%%%%%%%%%%%%%%%%%%%%%%%%%%%%%%%%%%%%%
\section{Introduction}
A neutrinoless double-beta decay ($0\nu2\beta$) is the key to addressing the mysteries of neutrinos: their light mass, mass hierarchy, and Majorana nature.
The observation of a $0\nu2\beta$ event would be direct proof of the Majorana-type particle nature of neutrinos~\cite{PhysRevD.25.774}. 
In the Majorana case, the mass of the left-handed neutrino might be small via the so-called See-Saw mechanism~\cite{SeeSaw}.
A hint of the neutrino mass hierarchy is given by neutrino-mass measuring experiments and the half-life of $0\nu2\beta$ ($T_{1/2}^{0\nu}$) with the relationship between the effective mass ($\langle m_{\beta\beta} \rangle$) and each neutrino mass ($m_1, m_2, m_3$) as
\begin{eqnarray}
    \left( T_{1/2}^{0\nu} \right)^{-1}
    &=& G^{0\nu} \left| M^{0\nu} \right|^2 \langle m_{\beta\beta} \rangle^2, \\
    \left| \langle m_{\beta\beta} \rangle \right| &\equiv&
    \left|
      \left| U_{e1}^{L}\right|^2 m_1 +
      \left| U_{e2}^{L}\right|^2 m_2 e^{i\phi_2} +
      \left| U_{e3}^{L}\right|^2 m_3 e^{i\phi_3}
    \right|,
\end{eqnarray}  
where $G^{0\nu}$ is a phase space factor, $M^{0\nu}$ is a nuclear matrix element, $e^{i\phi_{2,3}}$ are Majorana CP phases, and $U^L_{ej}$ ($j = 1-3$) is the neutrino mixing matrix.
The most strict lower limit on the half-life is $1.07 \times 10^{26}$~yr at 90\% C.L., reported by KamLAND (Kamioka Liquid scintillator AntiNeutrino Detector)-Zen~\cite{Gando2016}.

The sensitivity to $\langle m_{\beta\beta} \rangle$ is proportional to the square root of the exposure time with a background-free search, while it is to the fourth square root of the exposure time with some background~\cite{doi:10.1146/annurev.nucl.52.050102.090641}.
Hence, for the high-sensitivity $0\nu2\beta$ search, we require many target nuclei, a long livetime, and a background-free environment.
Unfortunately, environmental radioactive backgrounds exist at the Q-value of the $0\nu2\beta$, especially decay chains of uranium and thorium.

There are two ways to run and plan detectors for the high-sensitivity search.
One uses a high energy resolution and/or signal-tracking detector, such as CUORE, GERDA, MAJORANA, NEMO, and EXO~\cite{Alduino:2017ehq,Agostini2017,Aalseth2017,ARNOLD200579,PhysRevLett.120.072701}.
Such detectors have the strong advantage of less background contamination.
The high energy resolution enables us to use the narrow region of interest, which is less sensitive to the background. 
The tracking discrimination can determine the characteristic topology of the double-beta decay, which exhibits two beta tracks from a nucleus.
The other takes a large-volume liquid scintillator detector, such as KamLAND-Zen and SNO+~\cite{Gando2016,Andringa2015}.
They can use many target nuclei and collect high statistics; however, they do not have such tracking information nor such a good energy resolution.
Therefore, it is difficult for liquid scintillator detectors to discriminate the signal from the radioactive backgrounds.

From the KamLAND-Zen report~\cite{Gando2016}, the dominant background of the $0\nu2\beta$ search with ${}^{136}{\rm Xe}$ at the liquid scintillator detector is ${}^{214}{\rm Bi}$ daughter nuclei of the uranium series in and on the surface of the container.
The ${}^{214}{\rm Bi}$ has $Q_{\beta/\gamma}=3.2$~MeV with a broad spectrum over the 2.46-MeV Q-value of the ${}^{136}{\rm Xe}$ double-beta decay.
Usually, the ${}^{214}{\rm Bi}$ is identified with a 99.99\% efficiency in a liquid scintillator from the delayed-coincidence method using sequential events of ${}^{214}{\rm Bi}$-${}^{214}{\rm Po}$ within the time--space correlation.
However, the efficiency is small near the detector vessel because the delayed alpha-ray signal from ${}^{214}{\rm Po}$ is trapped at the non-scintillating vessel.
To suppress the contamination, the KamLAND-Zen experiment rejects events near the vessel within a region whose volume is almost half of the xenon-loaded liquid scintillator.

According to reference~\cite{Nakamura2011}, polyethylene naphthalate can be used as a scintillator with blue photon emission.
Generally, a polyethylene-naphthalate resin has a strong chemical resistance such that it is compatible with a strong solvent of a liquid scintillator, whereas a commercial plastic scintillator cannot be used.

This paper proposes the use of a polyethylene-naphthalate film scintillator as a liquid scintillator vessel (scintillation balloon) and presents feasibility studies for future KamLAND-Zen projects.

%%%%%%%%%%%%%%%%%%%%%%%%%%%%%%%%%%%%%%%%%%%%%%%%%%%%%%%%%%%%%%%%%%%%%%%%%%%%%%%%%%%%%%
\section{Requirements for Scintillation Balloon}
KamLAND is a 1-kton ultrapure liquid scintillator detector located 1,000-m underground from the top of Mt.~Ikenoyama, Japan \cite{Eguchi}.
Scintillation light is detected by the 1,879 photomultiplier tubes.
From the time--charge correlation of the photomultiplier tubes, event energy and vertex are reconstructed.
For a $0\nu2\beta$ decay search with ${}^{136}{\rm Xe}$, an inner balloon is installed to hold the xenon-loaded liquid scintillator at the center of the detector.
Each inner balloon of KamLAND-Zen~400 and KamLAND-Zen~800 \cite{Gando2019} was fabricated with a transparent nylon film thickness of 25-$\mu$m.
The liquid scintillator consists of 82\% decane, 18\% pseudocumene by volume, and 2.7~g/L of the fluor PPO (2,5-diphenyloxazole).

In order to use polyethylene-naphthalate film for a scintillation balloon with the liquid scintillator detector, we require a high transparency for the scintillation photons of the liquid scintillator, a sufficient scintillation light yield, chemical compatibility, and usability for the fabrication of the balloon shape.

% Transparency and emission spectra
\subsection{Transparency and Emission Spectra}
The balloon film must have a transparency larger than 90\% and no adsorption for event reconstruction at the emission wavelength of the liquid scintillator or wavelength shifter. 
The current xenon-loaded liquid scintillator used in KamLAND-Zen has a wavelength emission of 380~nm. 
Using bis-MSB, which is a well-known wavelength shifter for liquid scintillator detectors, the emission peak can be shifted to 420~nm.

% Light yield
\subsection{Light Yield}
To tag the sequential events of ${}^{214}{\rm Bi}$-${}^{214}{\rm Po}$ with the delayed-coincidence measurements, the light yield of the scintillation balloon should be sufficiently bright to observe a ${}^{214}{\rm Po}$ event.
The Q-value of an alpha-ray from ${}^{214}{\rm Po}$ is 7.7~MeV.
However, the ionization quenching effect changes the visible energy to 614~keV in the KamLAND liquid scintillator (hereinafter, referred to as ``${\rm keV_{KL}}$''), while the detector energy threshold is about 300~${\rm keV_{KL}}$. 
The visible energy ${\rm keV_{KL}}$ is defined as 2.22~MeV for a 2.22~MeV gamma-ray from a neutron-capture event on a proton in KamLAND and is proportional to the photon yield.

Therefore, the effective light yield from the scintillation balloon must be larger than at least 300~${\rm keV_{KL}}$, including the quenching effect, for the ${}^{214}{\rm Po}$ alpha-ray.
Its required light yield corresponds to about 2,100-photon emission from the scintilllation balloon, based on the 8,000-photons/MeV emission of the KamLAND liquid scintillator~\cite{KLpaper} and the cherenkov/scintillation photon yield ratio and the Birk's constant.

% Chemical resistance
\subsection{Chemical Compatibility}
The pseudocumene, a strong solvent in the liquid scintillator, undergoes chemical reactions with many organic plastics.
To use the vessel as the liquid scintillator, the scintillation balloon material must not dissolve, enabling long experiments to be conducted.

% Fabrication
\subsection{Fabrication}
\begin{figure}
  \centering
  \includegraphics[width=0.8\linewidth]{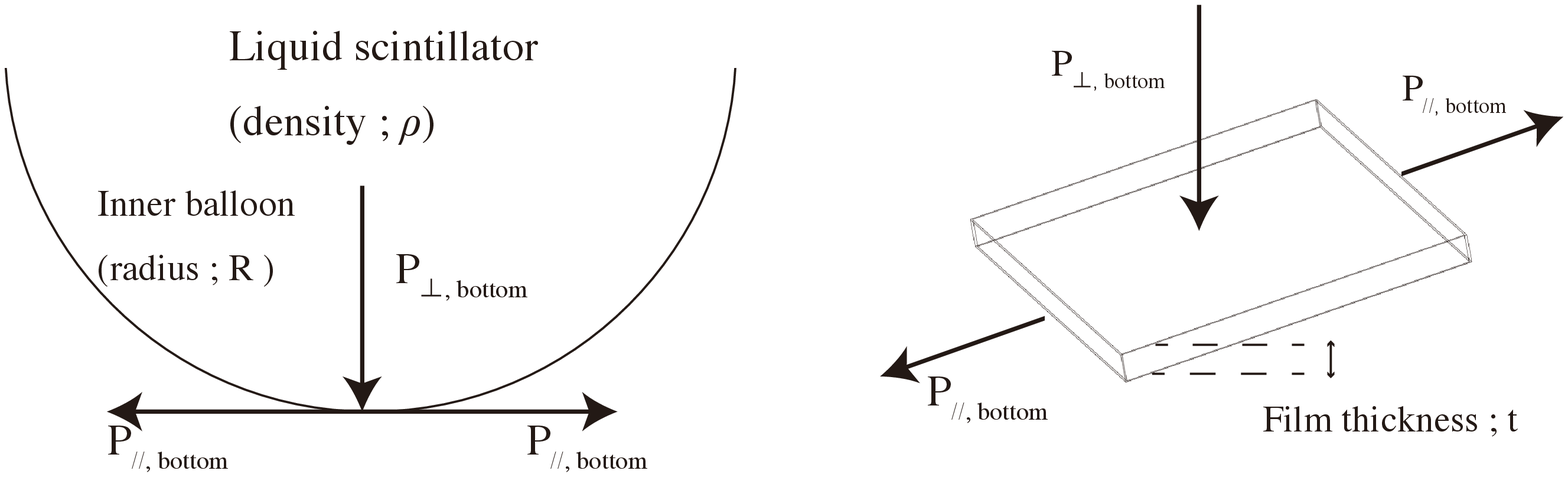}
  \caption{Schematic of load vectors at bottom of the balloon. The vertical pressure on the balloon surface from inside to outside depends on the density difference in/out of the liquid scintillator. A tension of the film, $P_{\parallel,\ {\rm bottom}}$, is numerically calculated.}
  \label{fig:BalloonLoadVector}
\end{figure}

It is difficult to use monolithic injection-type molding to construct the balloon for the liquid scintillator vessel because of the large detector diameter of about $3-4$~m.
The scintillation balloon will be fabricated by welding with gore-shaped films, identical to the current nylon-based inner balloon of KamLAND-Zen.
The breaking strength of scintillation balloons must be higher than the maximum load value, as shown in the bottom of Figure~\ref{fig:BalloonLoadVector}.
The maximum strain on the balloon bottom is
\begin{equation}
  P_{\perp,\ {\rm bottom}} = \rho  g  (H + R) ,
\end{equation}
where $\rho$ is the density difference in/out of the balloon, $g$ is the gravity acceleration constant, $H$ is the height of the top of the balloon from the center, and $R$ is the balloon radius.
The required strength can be described as
\begin{eqnarray}
  P_{\parallel,\ {\rm bottom}} &=& \frac{1}{2} \frac{R P_{\perp,\ {\rm bottom}}}{t} \\
  &=& \frac{1}{2} \frac{R(\rho g (H + R))}{t},
\end{eqnarray}
where $t$ is the thickness of the balloon film.
Considering an inner balloon radius of 2~m for a future phase of KamLAND-Zen, we require a film material strength of 4~MPa, corresponding to a strength of 1~N for a width of 1~cm and film thickness of 25-$\mu$m. 
In an actual case, considering the deformation of the balloon shape, a strength of 40~MPa with a safety factor of 10 is required.

% Radioactivity
\subsection{Radioactivity}
The radioactivity must be considered in rare decay search experiments.
Assuming a $\mathcal{O}(10^{-12})$~g/g contamination of ${}^{238}{\rm U}$/${}^{232}{\rm Th}$ and $\mathcal{O}(10^{-11})$~g/g of ${}^{40}{\rm K}$, the expected number of backgrounds from the radioactivity can be suppressed to a comparable amount of the ${}^{8}{\rm B}$ solar neutrino background, which is an unavoidable electron-scattering background for liquid-scintillator-type $0\nu2\beta$ search experiments.

\subsection{Material Candidates}
A nylon film satisfies our physical requirements of transparency and fabrication feasibility but exhibits no scintillation for ${}^{214}{\rm Po}$ alpha-rays.
An organic plastic scintillator material such as polystyrene exhibits scintillation but not chemical compatibility with a liquid scintillator.
Polyethylene-telepthalate is one of the candidates for the scintillation balloon and could be used as the scintillator with a wavelength emission of about 300~nm, chemical resistance, and high transparency. 
However, its light yield is small and emission wavelength too short for our use.

In this study, we focus on a polyethylene-naphthalate film, which has been recently reported as a new potential scintillator material.

%%%%%%%%%%%%%%%%%%%%%%%%%%%%%%%%%%%%%%%%%%%%%%%%%%%%%%%%%%%%%%%%%%%%%%%%%%%%%%%%%%%%%%
\section{Evaluation and Feasibility Study}
In this study, we use a polyethylene-naphthalate film as a candidate material for the scintillation balloon.
This section describes some feasibility studies.

\subsection{Emission Spectra and Transparency}
\begin{figure}
  \begin{center}
    \includegraphics[width=0.8\linewidth]{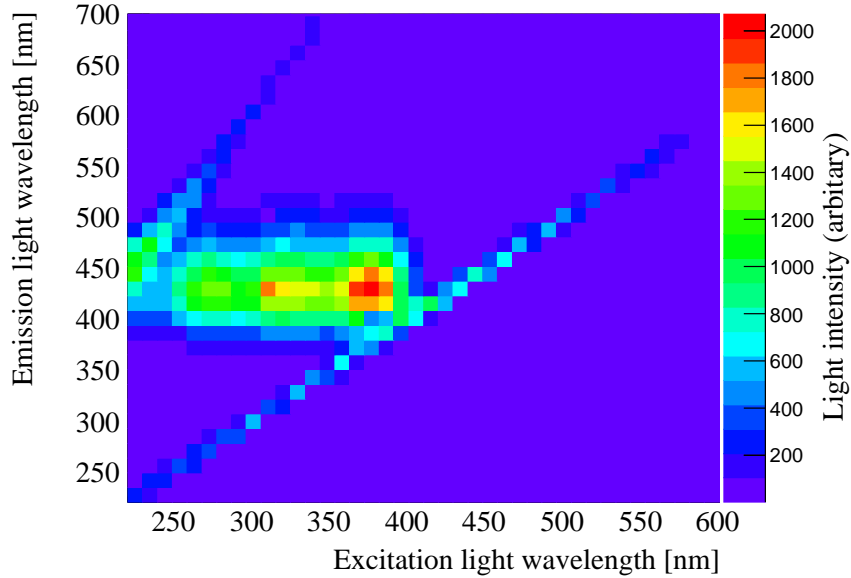}
    \caption{Emission wavelength of 2D plot for the polyethylene-naphthalate film.
    The $x$-axis and $y$-axis indicate the wavelength of excitation light and emission light, respectively.
    The $z$-axis color represents the relative light intensity. 
    This shows a $300-380$-nm absorption wavelength spectrum and $400-430$-nm emission wavelength spectrum. 
    The line from $({\rm EX}, {\rm EM}) = (250, 250)$ to $(550, 550)$ is the reflective spectrum on the film surface from the inputted excitation light, where EX is the excitation wavelength on the $x$-axis and EM is the emission wavelength on the $y$-axis. 
    The line through $({\rm EX}, {\rm EM}) = (200, 500)$ to $(350, 700)$ corresponds to the secondary reflective spectrum.}
    \label{ResultWavelength}
  \end{center}
\end{figure}

\begin{figure} 
  \centering
    \includegraphics[width=0.8\linewidth]{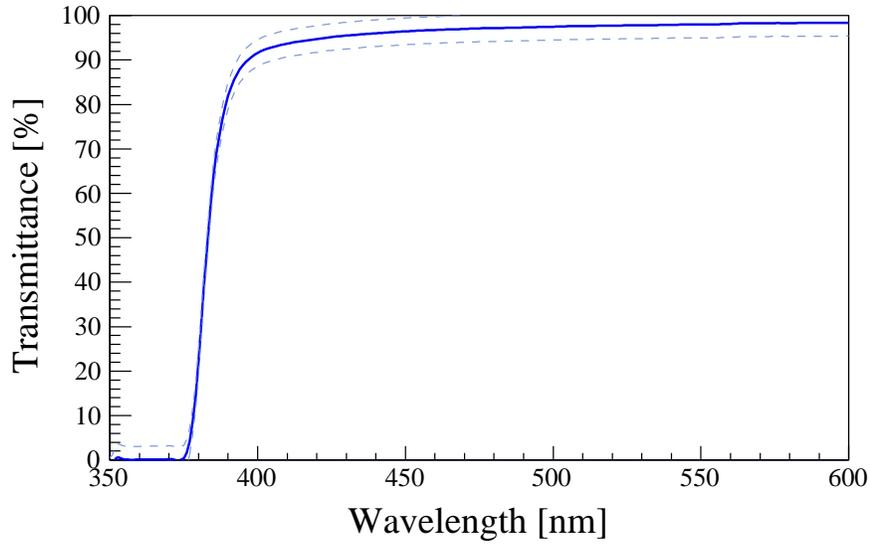}
    \caption{Transparency of a polyethylene-naphthalate film with a thickness of 50~$\mu$m (solid line). 
    The dashed lines indicate the 5\% range of systematic error.}
    \label{FigureTransparencyOfHighTransScintirex}
\end{figure}

The emission and absorption spectra were measured with the Hitachi F-2000 spectrofluorometer, as shown in Figure~\ref{ResultWavelength}.
The absorption peak is observed at about $300-380$~nm and the emission peak at about $400-430$~nm.
The transparency of the 50-$\mu$m-thick polyethylene-naphthalate film was measured with the Hitachi U-3900H spectrophotometer, as shown in Figure~\ref{FigureTransparencyOfHighTransScintirex}.
The result satisfies the 90\% transparency requirement at wavelengths $>$ 400~nm.
A low transparency range at wavelengths $<$ 380~nm corresponds to an absorption peak of polyethylene naphthalate, at which the current liquid scintillator has an emission peak.
The bis-MSB can shift the emission peak of the liquid scintillator to higher than the adsorption peak of the polyethylene naphthalate.

\subsection{Light Yield}
In order to identify the ${}^{214}{\rm Bi}$ background via ${}^{214}{\rm Bi}$-${}^{214}{\rm Po}$ delayed-coincidence measurement, the light yield of the scintillation balloon for an alpha-ray from ${}^{214}{\rm Po}$ should be brighter than 300~${\rm keV_{KL}}$ including the quenching effect.
We used ${}^{222}{\rm Rn}$--air, which is a parent nucleus for ${}^{214}{\rm Bi}$-${}^{214}{\rm Po}$.
Rn--air was dissolved into mineral oil in a vial, and eight layers of 50-$\mu$m-thick polyethylene-naphthalate films (total thickness of 400-$\mu$m) were placed.
The layered films were sufficient for an energy deposition of 3~MeV beta-rays from ${}^{214}{\rm Bi}$ to observe the scintillation photons.
The vial was placed on a 5.08-cm (2-in.) photomultiplier tube.
The delayed-coincidence events were selected within 1,000~$\mu$s between a prompt event of ${}^{214}{\rm Bi}$ and a delayed event of ${}^{214}{\rm Po}$.

\begin{figure}
  \centering    
    \includegraphics[width=0.7\linewidth]{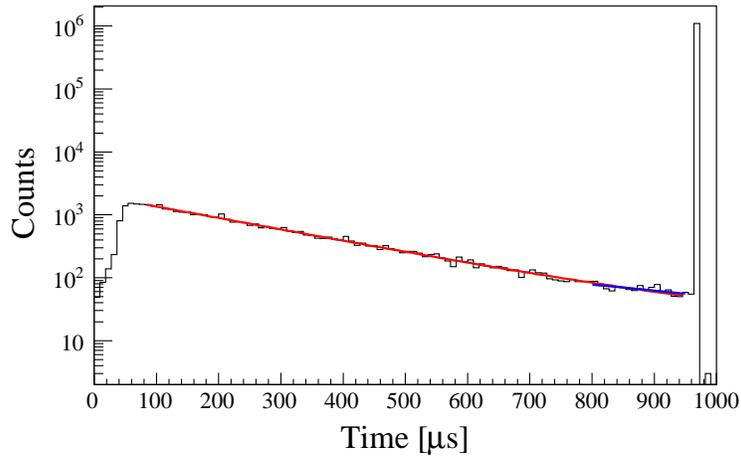}
    \caption{Time difference between prompt and delayed event for the delayed-coincidence measurement of ${}^{222}$Rn induced from a polyethylene-naphthalate film. 
    Before 100~$\mu$s, there is a dead time effect from the last event, leading to a missed event selection. 
    After 950~$\mu$s, a forced time-out signal appears.
    Lifetime fitting with the range of $100-950$ $\mu$s results in a half-life a 162 $\pm$ 2~$\mu$s, which is consistent with the half-life of ${}^{214}$Po as 164.3~$\mu$s.}
    \label{TimeDiffBiPoPEN}
\end{figure}

\begin{figure}
  \centering    
    \includegraphics[width=0.7\linewidth]{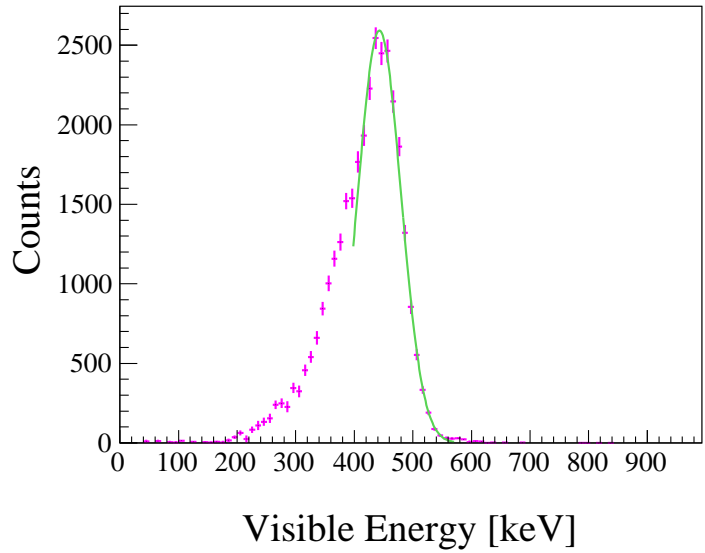}
    \caption{Light intensity of polyethylene-naphthalate film for ${}^{214}$Po event with the delayed-coincidence method. 
    The visible energy corresponds to 442.9 $\pm$ 0.7~${\rm keV_{KL}}$ as a result of one-side Gaussian fitting, except for the escape alpha-ray effect.}
    \label{BiPoLightIntensityPEN}
\end{figure}

The obtained time correlation is shown in Figure~\ref{TimeDiffBiPoPEN}, and an exponential fitting results in a half-life of 162 $\pm$ 2~$\mu$s, which is consistent with the half-life of 164.3~$\mu$s.
The delayed energy spectrum has a peak of 442.9 $\pm$ 0.7~${\rm keV_{KL}}$ based on the KamLAND-LS visible energy as shown in Figure \ref{BiPoLightIntensityPEN}.
This delayed energy of polyethylene naphthalate is sufficiently bright for the detector threshold.

\subsection{Chemical Compatibility}
\begin{figure}
  \centering
    \includegraphics[width=0.8\linewidth]{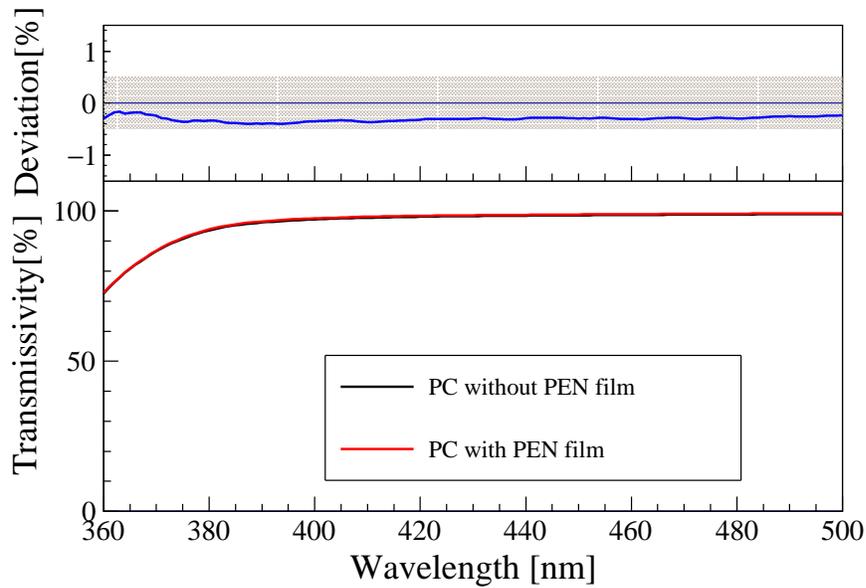}
    \caption{Transparency of pseudocumene with/without a polyethylene-naphthalate film after 2 months at 45~degC. 
    PC represents pseudocumene and PEN represents polyethylene naphthalate.
    The deviation is defined as the transparency difference of pseudocumene {\it with} a polyethylene-naphthalate film from {\it without} a polyethylene-naphthalate film. 
    They are consistent within the error range of 0.5\% (gray hatched region on the top figure).}
    \label{FigureTransparencyOfPseudocumeneWithPEN}
\end{figure}

We prepared two vials filled with 120~mL of pure pseudocumene.
One vial had pseudocumene only and the other had a piece of polyethylene-naphthalate film (25~mm $\times$ 25~mm square and 1~mm in thickness) corresponding to a 7.0-g/L contribution of polyethylene naphthalate, testing the tolerance to pseudocumene.
The vials were stored for 2 months at 45~degC, which corresponds to 12~months of tolerance testing at 15~degC based on the Arrhenius equation.
We compared the transparency of each pseudocumene with the Hitachi U-3900 spectrophotometer.
If polyethylene naphthalate dissolves into pseudocumene, the transparency will degrade.
The result of the difference between the two vials with/without polyethylene naphthalate is presented in Figure~\ref{FigureTransparencyOfPseudocumeneWithPEN}. 
It shows no dissolution within an error range of 0.5\%.

\subsection{Fabrication}
\begin{figure}
  \centering   
    \includegraphics[width=0.7\linewidth]{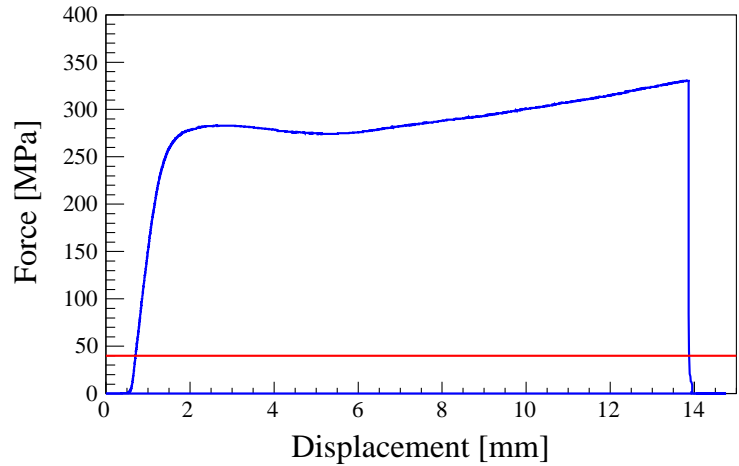}
    \caption{Film strength of 50-$\mu$m-thick polyethylene-naphthalate, and the required value of 40~MPa (red line). 
    We cut the film into a 1-cm-wide and 3-cm-long piece and strained its longer side.
    The $x$-axis indicates the displacement of the film length in the strain measurement, and the $y$-axis shows its measured power. 
    This shows the yield point at about ($x$, $y$) = (2~mm, 280~MPa) and the breaking point at about ($x$, $y$) = (14~mm, 330~MPa).}
    \label{FigureFilmStrengthResult}
\end{figure}

The breaking strain of the 50-$\mu$m-thick polyethylene-naphthalate film was measured with the IMADA ZTA-500 digital force gauge.
We cut the sample into a 1-cm-wide and 3-cm-long piece and strained its longer side.
Figure~\ref{FigureFilmStrengthResult} shows the relationship between the forced power and the displacement of the film.
The polyethylene-naphthalate film has a fracture point at 330~MPa.
This value satisfies our requirement of 40~MPa.

\begin{figure}
  \begin{minipage}{0.48\hsize}
    \centering
    \includegraphics[width=0.7\linewidth]{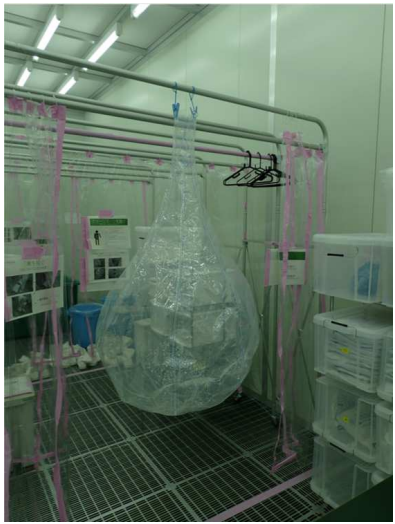}
  \end{minipage}
  \begin{minipage}{0.48\hsize}
    \centering
    \includegraphics[width=0.7\linewidth]{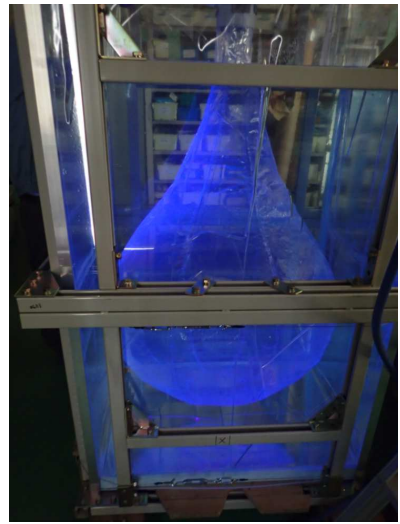}
  \end{minipage}
  \caption{Test-size scintillation balloon in a clean room (left) and 
water tank with ultraviolet light (right). 
  The balloon in the left figure is filled with air, while that in the right figure is filled with water.}
  \label{TestSizeScintillationBalloon}
\end{figure}

As a test for fabricating the inner balloon with the polyethylene-naphthalate film, we produced a test-sized scintillation balloon with an 800-mm diameter and 50-$\mu$m-thick from commercial polyethylene-naphthalate film roll.
We cut the polyethylene naphthalate into eight boat-shaped gores.
Because of the strong chemical compatibility of polyethylene naphthalate, it is generally difficult to bond with adhesives.
We chose heat-welding with the following tuned parameters: 237~degC as the maximum temperature for heating, 3.5~s as the heating time, and 80~degC as the cooling temperature.
Figure~\ref{TestSizeScintillationBalloon} shows the fabricated test-size scintillation balloon in a clean room (left) and in a water tank with ultraviolet light (right).  
We established methods to make a balloon with the polyethylene-naphthalate film.

\subsection{Radioactivity}
Radioactive impurities in the polyethylene-naphthalate film were measured with ICP-MS (Agilent Technologies 7500) for ${}^{238}{\rm U}$ and ${}^{232}{\rm Th}$ and frame-spectrophotometry (Varian SpectrAA-55B) for ${}^{\rm nat}{\rm K}$ by the NTT Advanced Technology Corporation.
The contaminations of ${}^{238}{\rm U}$, ${}^{232}{\rm Th}$, and ${}^{\rm nat}{\rm K}$ in a polyethylene-naphthalate film were $3.6 \times 10^{-11}$~g/g, less than $5 \times 10^{-12}$~g/g, and $2.0 \times 10^{-10}$~g/g ($0.02 \times 10^{-12}$~g/g for ${}^{40}{\rm K}$ assuming natural abundance), respectively. 
This indicates that the impurities of ${}^{232}{\rm Th}$ and ${}^{40}{\rm K}$ are satisfied with our requirement.
Although the ${}^{238}{\rm U}$ contamination is one order higher than our requirement, it can be suppressed by the ${}^{214}{\rm Bi}$-${}^{214}{\rm Po}$ tagging, as described in the next section.

%%%%%%%%%%%%%%%%%%%%%%%%%%%%%%%%%%%%%%%%%%%%%%%%%%%%%%%%%%%%%%%%%%%%%%%%%%%%%%%%%%%%%%
\section{Discussion and Summary}
We evaluated a balloon-shaped vessel made of a polyethylene-naphthalate film to search for a rare decay using a liquid scintillator detector in terms of organic solvent tolerance, high transparency at wavelengths longer than 400~nm, and availability of heat-welding.
Although the ${}^{238}{\rm U}$ contamination is one order higher than our requirement, it will be acceptable if the scintillation balloon identifies the ${}^{214}{\rm Bi}$ background with a 90\% efficiency.
Actually, the KamLAND trigger with the 300-${\rm keV_{KL}}$ threshold can detect the alpha-rays of ${}^{214}{\rm Po}$ with an energy of $442.9 \pm 0.7$~${\rm keV_{KL}}$ with a 99.7\% efficiency.
Therefore, the one-order-higher radioactivity of uranium is negligible for the KamLAND-Zen by ${}^{214}{\rm Bi}$ - ${}^{214}{\rm Po}$ tagging.

The current KamLAND liquid scintillator does not include bis-MSB, and the emission peak of the liquid scintillator is about 380~nm while polyethylene naphthalate has a similar absorption peak.
It is difficult to use polyethylene naphthalate as a scintillation balloon for the current phase of KamLAND-Zen regarding re-emission scintillation.
Hence, we propose this scintillation balloon for a future project of KamLAND2-Zen using the new linear-alkyl-benzene-based liquid scintillator with bis-MSB or other experiments that use an organic liquid scintillator to search for a rare decay.

%%%%%%%%%%%%%%%%%%%%%%%%%%%%%%%%%%%%%%%%%%%%%%%%%%%%%%%%%%%%%%%%%%%%%%%%%%%%%%%%%%%%%%
\section*{Acknowledgment}
This work was supported by JSPS KAKENHI Grant Numbers 25707014, 26287035, and 15J02193.
We are grateful to TEIJIN for providing us with polyethylene-naphthalate film samples.
We also acknowledge technical staff of the Research Center for Neutrino Science, Tohoku University.
We would like to thank Editage (www.editage.jp) for English language editing.

%\bibliographystyle{ptephy.bst}
%\bibliography{ScintillationBalloon}

%\bibliographystyle{ptephy.bst}
%\bibliography{dummy}

%%%%%%%%%%%%%%%%%%%%%%%%%%%%%%%%%%%%%%%%%%%%%%%%%%%%%%%%%%%%%%%%%%%%%%%%%%%%%%%%%%%%%%

%%%%%%%%%%%%%%%%%%%%%%%%%%%%%%%%%%%%%%%%%%%%%%%%%%%%%%%%%%%%%%%%%%%%%%%%%%%%%%%%%%%%%%

%%%%%%%%%%%%%%%%%%%%%%%%%%%%%%%%%%%%%%%%%%%%%%%%%%%%%%%%%%%%%%%%%%%%%%%%%%%%%%%%%%%%%%
%figures
\clearpage

\end{document}